\documentclass {article}
\usepackage{fleqn,espcrc2,epsf}

\newcommand{\be}{\begin{equation}}
\newcommand{\ee}{\end{equation}}

\newcommand{\order}{{\cal O}}

\newcommand{\AmS}{{\protect\the\textfont2
  A\kern-.1667em\lower.5ex\hbox{M}\kern-.125emS}}

\title{Matching of the Heavy-Light Currents with NRQCD Heavy and Improved Naive Light Quarks}

\author{E.Gulez\address{Physics Department, The Ohio State
        University, Columbus, OH 43210, USA.},
        J.Shigemitsu$^{\rm a}$,
        M.Wingate\address{Institute for Nuclear Theory, University of
        Washington, Seattle, WA 98195, USA.}
              }

\begin{document}

\begin{abstract}
One-loop matching of heavy-light currents is carried out for a
highly improved lattice action, including the effects of mixings
with dimension $4$ $\cal{O}$$(1/M)$ and $\cal{O}$$(a)$
operators. We use the NRQCD action for heavy quarks, the Asqtad
improved naive action for light quarks, and the Symanzik improved
glue action. These results are being used in recent heavy meson decay constant and semileptonic form factor calculations on the MILC dynamical configurations.
\vspace{1pc}
\end{abstract}


\maketitle

\section{Introduction}
\begin{figure}
\epsfxsize=7.5cm
\centerline{\epsfbox{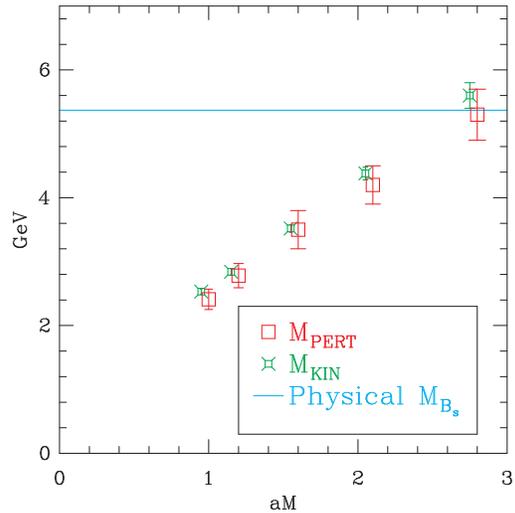}}
\caption{Comparison of the kinetic and perturbative meson masses. The rightmost point at $aM_0=2.8$ corresponds to the physical $B_s$ meson. The data for this plot is from \cite{wingate}. The points representing $M_{KIN}$ were shifted to the left for clarity.}
\end{figure}
Matching of the heavy-light currents between lattice simulations and
full continuum theory is important for studies of heavy meson
leptonic and semileptonic decays. As statistical and discretization errors are reduced, chiral
extrapolation and operator matching errors dominate, and more accurate
matching calculations become necessary.
As part of the work of studying heavy mesons, one-loop
perturbative matching of the heavy-light currents was performed. In this talk we focus on the matching of the temporal currents $V_0$ and $A_0$. Our calculations are correct through
$\order (\alpha_s,a\alpha_s,\alpha_s/aM,\alpha_s\Lambda_{QCD}/M)$. Matching of the spatial components $V_k$ and $A_k$ to the same level of accuracy has now also been completed.

\section{Actions Used and Calculational Strategies}
The actions used are highly improved lattice actions. The AsqTad improved naive light quark action, which is correct through $\cal{O}($$a^2$$)$ was used. For the heavy quarks, we used an improved NRQCD action. This action is correct through $\cal O$$ (1/M^2,a^2)$, and is the same action that is being used in numerical simulations on MILC dynamical configurations. For the gluon, the Symanzik improved glue action was used.

Two independent methods were employed for the matching calculation:First, Mathematica was used to do the Dirac algebra and where necessary, to take derivatives with respect to external momenta. The result was converted into Fortran format and numerical integration was done with the Fortran Vegas program.
Second, a C++ code originally developed for a similar calculation with other types of actions by C. Morningstar was modified and used. The derivatives were taken by automatic differentiation.

\begin{figure}
\epsfxsize=7.5cm
\centerline{\epsfbox{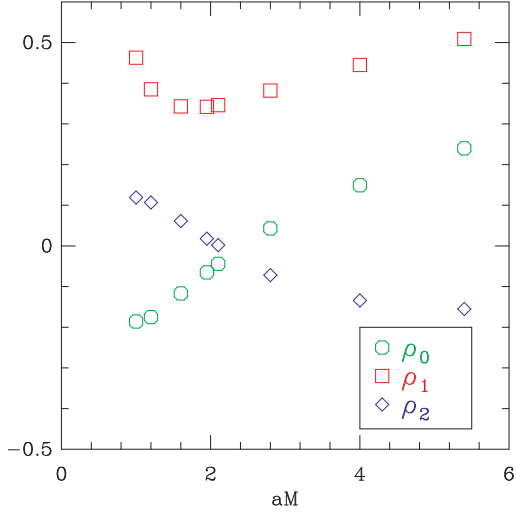}}
\caption{Some results for the perturbative matching coefficients $\rho_j$. The data for this plot is from \cite{ours}.}
\end{figure}

\section{Self Energy}
For massless Asqtad Quarks:
\begin{eqnarray}
Z_q &=& 1 + \alpha_s \, [\,C^{IR}_q  \; + \; C_q\,] \; + \; \order (\alpha_s^2) \nonumber \\
C^{IR}_q &=& \frac{1}{3 \, \pi} \left[1 + (\xi - 1) \right] \;
{\rm ln}( a^2 \lambda^2) \nonumber \\
C_q &=& C_q^{reg} \; + \; C_q^{tad} \; + \; C_q^{u0} \nonumber \\
C_q^{u0} &=& \; - \; \left[\,4 - \frac{1}{4} - \frac{3}{2}\right] \,
 u^{(2)}_0 = \; - \; \frac{9}{4} \; u^{(2)}_0 \, ,\nonumber \\
u_0 &\equiv& 1 - \alpha_s \, u^{(2)}_0 + \order (\alpha_s^2) \, .\nonumber 
\end{eqnarray}
\begin{figure}
\epsfxsize=7.5cm
\centerline{\epsfbox{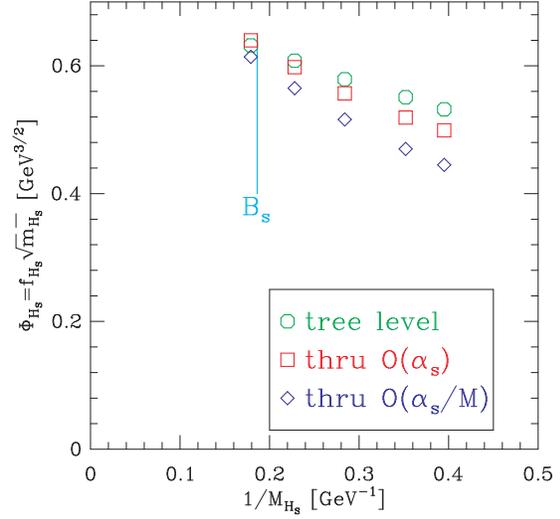}}
\caption{The effects of matching on $f_{H_s}\sqrt{M_{H_s}}$. At the value of the inverse $B_s$ mass, the effect of the perturbative correction is very small (within 3\%). The data for this plot is in \cite{wingate}.}
\end{figure}

$u_0^{(2)} = 0.767$ for
the plaquette definition of $u_0$ and $u_0^{(2)} = 0.750$ for
the Landau link definition, and $\lambda$ is the gluon mass.

For heavy Quarks:
\begin{eqnarray}
Z_Q &=& 1 \; + \; \alpha_s \, [\, C_Q^{IR} \; + \; C_Q \,] \; + \;
\order (\alpha_s^2)  \nonumber \\
Z_M &=& 1 \; + \; \alpha_s \,  C_M   \; + \;
\order (\alpha_s^2)  \nonumber \\
a \, E_0 &=& \qquad \; \alpha_s \,  C_{E0}   \; + \;
 \order (\alpha_s^2)\nonumber \\
C^{IR}_Q &=& \frac{1}{3 \, \pi} \left[-2 + (\xi - 1) \right] \;
{\rm ln}( a^2 \lambda^2) \,\nonumber 
\end{eqnarray}
$Z_q$ was previously calculated by
H. Trottier using twisted boundary conditions
as IR regulator. Our results for the 
lattice-continuum difference are in agreement.
\section{Kinetic and Perturbative Masses for Heavy-Light Mesons }
Kinetic mass is obtained nonperturbatively from finite momentum correlators using the meson dispersion relation;
\begin{equation}
M_{KIN} = \frac{|p|^2-\Delta^2}{2\Delta}\nonumber 
\end{equation}
where $\Delta=E(p)-E(0)$.
The perturbative mass is estimated using the zero momentum correlator and the one-loop heavy quark mass renormalization
\begin{equation}
M_{PERT}=E_{SIM}(0)+Z_MM_0-E_0\nonumber 
\end{equation}
As shown in Fig. 1, the results are in agreement with each other.
\section{Current Matching}
The lattice currents that are matched with the corresponding continuum ones are listed in \cite{morningstar} The continuum currents are a linear combination of lattice ones:

\begin{eqnarray}
\langle A_\mu \rangle _{CONT}&=&\sum_j{C^{A_\mu}_j}\langle J^j_{A_\mu} \rangle _{LAT},\nonumber \\
\langle V_\mu \rangle _{CONT}&=&\sum_j{C^{V_\mu}_j}\langle J^j_{V_\mu} \rangle _{LAT}\nonumber 
\end{eqnarray}
The matching coefficients for the axial vector and vector coefficients are the same due to the good chiral properties of the AsqTad action;  $C^{A_\mu}_j=C^{V_\mu}_j$. The perturbative expansion of the matching coefficients goes as,
\begin{eqnarray}
C_{0,1}&=&1+\alpha\rho_{0,1}+\order (\alpha^2),\nonumber  \\
C_{2}&=&\alpha\rho_{2}+\order (\alpha^2),\nonumber
\end{eqnarray}
The goal is to find the perturbative matching coefficients $\rho_j$.  As seen in Fig. 2, these matching coefficients are small and well behaved, which shows that perturbation theory is working well. For the numerical values of the coefficients, see \cite{ours}.As an example, the effect of the matching on $\Phi_{H_s}=f_{H_s}\sqrt{M_{H_s}}$ is plotted in Fig 3. The difference is within 3\% around the physical $B_s$.

\section{Summary}

We have performed one-loop matching of 
the lattice NRQCD+AsqTad heavy-light 
current to the corresponding continuum current for 
different heavy quark masses. The calculation is correct through $\order (\alpha_s/aM)$. The perturbative matching coefficients are 
found to be small and well behaved. These matching coefficients are important for the study of leptonic and semileptonic decays of $B$ and $D$ mesons and of neutral $B$ mixing. The results of this work is used for calculating the decay constants for $B_s$, $D_s$ \cite{wingate} and $B$ \cite{alan}. These results are also important ingredients in calculation of form factors $f_0$, $f_+$ \cite{shigemitsu}. Similar perturbative matching for four fermion operators is underway.  

\section{Acknowledgments}
This work was supported by the DOE.

\end{document}